\documentclass[11pt,a4paper,nofootinbib,superscriptaddress]{revtex4-1}

\pdfoutput=1
\usepackage[utf8]{inputenc}
\usepackage{graphicx,bm,bbm}
\usepackage{amssymb,graphicx,epstopdf}
\usepackage{slashed,subfigure}
\usepackage{caption}
\usepackage{xcolor}
\usepackage{amsmath,mathtools}
\usepackage{epstopdf,dcolumn}
\usepackage{soul}
\usepackage{mathtools}
\allowdisplaybreaks
\usepackage[normalem]{ulem}
\usepackage{cancel}
\usepackage{xcolor}
\usepackage[colorlinks=true,linktocpage=true,linkcolor=blue,citecolor=blue]{hyperref}
\usepackage{xcolor}
\usepackage{braket}
\allowdisplaybreaks
\usepackage{enumitem}
\usepackage{mathrsfs}

\def\be {\begin{equation}}
\def\ee {\end{equation}}
\def\bea {\begin{eqnarray}}
\def\eea {\end{eqnarray}}
\def\bc {\begin{center}}
	\def\ec {\end{center}}
\def\nn {\nonumber}
\def\eps {\epsilon}

\def\mn {\mu\nu}
\def\al {\alpha}
\def\om{\omega}
\def\({\left(}
\def\){\right)}
\def\[{\left[}
\def\]{\right]}

\DeclareGraphicsExtensions{.jpg,.pdf,.eps}

\begin{document}

	\title{Shear Viscosity of hadronic matter at finite temperature and magnetic field}
	
	\author{Ritesh Ghosh}
	\email{ritesh.ghosh@saha.ac.in}
	\affiliation{
		Theory Division, Saha Institute of Nuclear Physics, A CI of Homi Bhabha National Institute, 1/AF, Bidhannagar, Kolkata 700064, India}
	
	\author{Najmul Haque}
	\email{nhaque@niser.ac.in}
	\affiliation{School of Physical Sciences, National Institute of Science Education and Research, An OCC of Homi Bhabha National Institute,\\  Jatni, Khurda 752050, India}

	\begin{abstract}
		We calculate the transport coefficient of hadronic matter in the presence of temperature and magnetic field using the linear sigma model. In the relaxation time approximation, we estimate the shear viscosity over entropy density $\eta/s$. The point-like interaction rates of hadrons are evaluated through the $S$-matrix approach in the presence of a magnetic field to obtain the temperature and magnetic field-dependent relaxation time. We observe that the transport coefficients are anisotropic in the presence of the magnetic field. We calculate the temperature and magnetic field-dependent anisotropic shear viscosity coefficients by incorporating the estimated relaxation time. The value of viscosity over entropy density is lower in the presence of a magnetic field than the value of it in a thermal medium. The behavior of the perpendicular components of the shear-viscous coefficient is also discussed. We consider the temperature-dependent hadron masses from mean-field effects in this work. 
	\end{abstract}
	
	\maketitle 
	\newpage
	
	\section{Introduction}
	In relativistic heavy-ion collision experiments at the Large Hadron Collider (LHC) and Relativistic Heavy Ion Collider (RHIC), a novel state of quarks and gluons, i.e., quark-gluon plasma (QGP)~\cite{Muller:1983ed} is produced as a near-perfect fluid~\cite{ALICE:2010khr,STAR:2005gfr,PHENIX:2004vcz}. The elliptic flow~\cite{Romatschke:2007mq,Gale:2013da} data indicates the smallest viscosity to entropy density ratio $(\eta/s)$ of the QGP medium. The produced QGP medium shows the collective behavior, and it undergoes space-time evolution and finally emanates to the hadronic phase. The transverse momentum spectra and the collectivity of the produced particles can be studied from the hydrodynamical modeling~\cite{Gale:2013da}. The transport coefficients are used as the input parameters for the hydrodynamic simulations. 
	
	More research interests have grown in the non-central heavy-ion collisions through the last decade. Several studies~\cite{Kharzeev:2007jp, Skokov:2009qp} suggest that a strong magnetic field is produced in non-central heavy-ion collisions in the perpendicular direction to the reaction plane. Initially, at the time of the collisions, the magnitude of the produced magnetic field can be of the order $10^{18}$\,G at RHIC and $10^{19}$\,G at LHC~\cite{Voronyuk:2011jd,Deng:2012pc}. The magnitude of the produced magnetic field depends on several parameters such as impact parameter, the conductivity of the medium, collision energy, etc. The strong field created in the heavy-ion collisions (HIC) decreases sharply with time~\cite{Bzdak:2012fr}. However, some studies~\cite{Tuchin:2013bda,Tuchin:2013ie,Inghirami:2016iru} have proclaimed that the presence of finite electric conductivity of the medium can extend the lifetime of the magnetic field. In recent times, the various properties of hot and dense matter have been investigated in the presence of a finite magnetic field. Different novel phenomena like chiral magnetic effect (CME)~\cite{Kharzeev:2013ffa}, magnetic catalysis~\cite{Mueller:2015fka}, inverse magnetic catalysis at finite temperature, thermodynamic properties~\cite{Bandyopadhyay:2017cle,Karmakar:2019tdp}, properties of quarkonia~\cite{Singh:2017nfa, Ghosh:2022sxi}, dilepton production~\cite{Tuchin:2013bda,Das:2019nzv,Das:2021fma, Bandyopadhyay:2016fyd}, chiral susceptibility~\cite{Ghosh:2021knc}, photon damping rate~\cite{Ghosh:2019kmf} and so on have been studied over the last few years. In the presence of a magnetic field, magnetohydrodynamics (MHD) simulations~\cite{Hongo:2013cqa,Pu:2016ayh,Roy:2015kma} have been developed to describe the fluid dynamics of strongly interacting matter. In this context, the  transport coefficients are relevant quantities to study, namely, shear viscosity~\cite{Das:2019pqd,Tuchin:2011jw,Dey:2019vkn,Li:2017tgi}, bulk viscosity~\cite{Kurian:2018dbn,Hattori:2017qih,Kadam:2014xka} and electrical conductivity~\cite{Ghosh:2019ubc,Fukushima:2017lvb,Huang:2011dc,Das:2019pqd} in hadronic and quark matter in presence of the constant magnetic field. All the transport coefficients become anisotropic in the magnetic field, and one gets five coefficients of shear viscosity, three coefficients for conductivity, and two bulk viscosity coefficients~\cite{Huang:2011dc}.

	This article evaluates the shear viscosity coefficients of hadronic matter in a strong magnetic field using the linear sigma model (LSM). The LSM is one of the simple models to study the hadronic system and was first introduced by Gell-Mann and L{\'e}vy~\cite{Gell-Mann:1960mvl}. Several works have been done considering this as a low-energy effective model during the last few years as it mimics the low-energy QCD region. Recent attempts have extended the LSM by including quarks~\cite{Ayala:2018zat,Das:2019ehv} and vector mesons in this model~\cite{Divotgey:2016pst}. Chiral phase transition~\cite{Petropoulos:2004bt}, pion condensate~\cite{Loewe:2013coa} and neutral pion mass~\cite{Das:2019ehv} in the presence of the external magnetic field and so on have been studied using the LSM. In Refs.~\cite{Chakraborty:2010fr,Heffernan:2020zcf} the authors have calculated the transport coefficients of hadronic matter at finite temperature using the LSM. The results show that the shear viscosity to entropy density ratio $(\eta/s)$ has a minimum at the crossover temperature. In contrast, the bulk viscosity to entropy density ratio $(\xi/s)$ has a maximum at the crossover temperature. As a first attempt, we study the viscous shear coefficient of hadronic matter in a magnetic field for zero chemical potential in this present work. There are five shear-viscous coefficients in a nonzero magnetic field, and we have studied all the coefficients. In the presence of a magnetic field, the relaxation times are estimated through the S-matrix approach. In these calculations, we would get the expressions of the matrix elements in terms of the Landau level summation. Our study considers only the lowest Landau level (LLL) contribution.
	
	The paper is arranged as follows: In Sec.~\ref{sec:viscosity} we review the formalism for the estimation of the shear viscosity coefficients in the presence of the external magnetic field within relaxation time approximation. In Sec.~\ref{sec:LSM} we discuss the basics of the linear sigma model (LSM) and the thermodynamic quantities that are used in the calculations. We have calculated the scattering amplitude and interaction rate in the presence of the magnetic field In subsection~\ref{sec:scattering}. Expressions for the interaction rate in the pure thermal medium are also discussed here. Incorporating the interaction rate, we finally obtain the anisotropic shear viscous coefficients, and presented is the result Sec.~\ref{sec:result}. Finally, we summarize in section~\ref{sec:summary}.

	\section{Anisotropic Viscosity coefficients in non zero magnetic field}
	\label{sec:viscosity}
	We will study the transport properties of a hadronic medium in the presence of a magnetic field.
	Transport coefficients can be calculated using two popular approaches: kinetic theory~\cite{Gavin:1985ph} and Kubo framework~\cite{Jeon:1994if, Ghosh:2020wqx, Satapathy:2021cjp}. Here we follow the former approach and briefly discuss this formalism in the relaxation approximation (RTA) technique~\cite{Tuchin:2011jw, Ghosh:2018cxb}. In a magnetic field, the transport coefficients for the charged particles become anisotropic, whereas the neutral particles contribute to the isotropic coefficients only. 
	
	In a magnetic field, the Boltzmann equation for single hadron species is written as 
	\bea
	p^\mu \partial_\mu f_a+ q F^{\mn} p_\nu\, \frac{\partial f_a}{\partial p^{\mu}} = C[f_a],\label{boltzmann_def}
	\eea
	where $F^{\mn}$ is the electromagnetic field tensor. In absence of electric field, $F^{\mn}=-B \,b^{\mn}$ where $b^{\mn}=\eps^{\mn \al\beta}b_\al u_\beta$ with fluid four velocity $u^\mu$.  The unit vector $b^{\mu}$ is defined as $b^{\mu}=\frac{B^{\mu}}{B}$. $C[f_a]$ is known as the collision integral, $p^\mu$ is the four momenta of the particle and $q$ is the electric charge of the particle. 
	
	In relaxation time approximation (RTA), the Boltzmann equation~\ref{boltzmann_def} is given by
	\bea
	p^\mu \partial_\mu f_a+ q F_{\mn} p_\nu\, \frac{\partial f_a}{\partial p^{\mu}} = -\omega_a (u\cdot p)\,\delta f_a,
	\eea
	where $\omega_a$ is frequency of interaction defined as the inverse of the equilibration time i.e.
	\bea
	\omega_a(E)=\tau_a^{-1}(E).
	\eea
	Assuming that the system is meagerly out of equilibrium, we can write the distribution function as,
	\bea
	f_a(x,p)=f_a^0\left(1+\phi_a(x,p)\right)=f_a^0+\delta f_a.
	\eea 
	For a small deviation from equilibrium, we can write the Boltzmann equation as 
	\bea
	p^\mu \partial_\mu f_a^0=\left(-\frac{u\cdot p}{\tau_a}\right)\left(1-\frac{qB \tau_a}{u\cdot p} b^{\mn}p_\nu \frac{\partial}{\partial p^\mu}\right)\delta f_a,
	\label{Boltzmann_eq}
	\eea 
	where $f_a^0=\exp{(-u_\al p^\al/T)}$ is equilibrium distribution function.

	Now, in general, the energy-momentum tensor is written as 
	\bea
	T^{\mn}&=&T^{\mn}_0+\Delta T^{\mn},
	\eea
	where $T^{\mn}_0$ represents the energy-momentum tensor in local equilibrium and $\Delta T^{\mn}$ is the deviation from the equilibrium. $T^{\mn}$ is given as 
	\bea
	T^{\mn}&=& \sum_a\int \frac{d^3p}{(2\pi)^3}\frac{p_a^\mu p_a^\nu}{E_a} f_a+  \sum_a\frac{|qB|}{2\pi}\int \frac{dp_3}{2\pi}\frac{\tilde p_a^\mu \tilde p_a^\nu}{E_a} f_a,
	\eea
	where the sum is over uncharged and charged particles in the first and second term respectively. In the second term phase factor is modified in presence of strong magnetic field.

	For shear viscosity, in presence of magnetic field we can express $\delta f_a$ in terms of fourth rank projection tensors as 
	\bea
	\delta f_a&=&\sum_{m=-2}^{2} c^m C^{(m)}_{\mn\al\beta}\, p^\mu p^\nu V^{\al\beta},
	\label{del_f_a}
	\eea
	where $V^{\al\beta}=\frac{1}{2}(\partial^\al u^\beta+\partial^\beta u^\al)$. There are five complex coefficients $c_m$.
	
	Shear viscous tensor $(\pi^{\mn})$ is given as 
	\bea
	\pi^{\mn}=\eta^{\mn\al\beta} V_{\al\beta},
	\label{eta_mu_nu_al_beta}
	\eea
	where the general form of $\eta^{\mn\al\beta}$ in presence of the magnetic field could be expressed with the fourth rank projection tensors as
	\bea
	\eta^{\mn\mu'\nu'}&=&\sum_{m=-2}^2 c^m C^{(m)\mn\mu'\nu'}.\label{4-tensor_def}
	\eea
	Here we introduce three second-rank projection tensors given as
	\bea
	P^{(0)}_{\mn}&=& b_\mu b_\nu,\\
	P^{(1)}_{\mn}&=& \frac{1}{2}(\Delta_{\mn}-b_\mu b_\nu+ i b_{\mn}),\\
	P^{(-1)}_{\mn}&=& \frac{1}{2}(\Delta_{\mn}-b_\mu b_\nu- i b_{\mn}),
	\eea
	where $\Delta^{\mn}=g^{\mn}-u^\mu u^\nu$.
	Fourth-rank tensor defined in eq.~\eqref{4-tensor_def} can be written in terms of two second-rank 
	tensors as~\cite{Dash:2020vxk, Hess:2015szz}
	\bea
	\mathscr{P}^{(m)}_{\mn\al\beta}&=&\sum_{m_1=-1}^{1}\sum_{m_2=-1}^1 P^{(m_1)}_{\mu\al}P^{(m_2)}_{\nu \beta}\delta(m,m_1+m_2).
	\eea
	In terms of real coefficients eq.~\eqref{eta_mu_nu_al_beta} can be written as
	\bea
	\eta_{\mn\al\beta} &=& c^0 \mathscr{P}^{(0)}_{<\mn>\al\beta}\nn\\
	&+&\sum_{m=1}^{2}\left\{c^{m+}
	\left(\mathscr{P}^{(m)}_{<\mn>\al\beta}+\mathscr{P}^{(-m)}_{<\mn>\al\beta}\right)+i c^{m-}
	\left(\mathscr{P}^{(m)}_{<\mn>\al\beta}-\mathscr{P}^{(-m)}_{<\mn>\al\beta}\right)\right\},
	\eea
	where $\mathscr{P}^{(m)}_{<\mn>\al \beta}=
	\mathscr{P}^{(m)}_{\mn\al\beta}+\mathscr{P}^{(m)}_{\nu\mu\al\beta}$.
	The coefficients $c^{m+}$ and $c^{m-}$ are real and imaginary parts of coefficients $c^m$. Three coefficients $c^0$, 
	$c^{1+}$ and $c^{2+}$ are even functions of the magnetic field whereas other two coefficients $c^{1-}$ and $c^{2-}$ are odd functions
	of magnetic field. The shear viscosity obeys the condition $\eta^{\mn\mu'\nu'}(B^\al)=\eta^{\mn\mu'\nu'}(-B^\al)$, the symmetry principle for transport coefficients.
	
	Now we can represent eq.~\eqref{del_f_a} with real coefficients i.e.
	\bea
	\delta f_a &=& \bigg[c^0 \mathscr{P}^{(0)}_{<\mn>\al\beta}\nn\\
	&+&\sum_{m=1}^{2}\left\{c^{m+}
	\left(\mathscr{P}^{(m)}_{<\mn>\al\beta}+\mathscr{P}^{(-m)}_{<\mn>\al\beta}\right)+i c^{m-}
	\left(\mathscr{P}^{(m)}_{<\mn>\al\beta}-\mathscr{P}^{(-m)}_{<\mn>\al\beta}\right)\right\}\bigg]p^\mu p^\nu V^{\al \beta}.
	\eea
	In integral form we can write the shear viscous tensor as 
	\bea
	\pi^{\mn}&=&\frac{1}{15} \sum_a\sum_{m=-2}^{2} \int \frac{d^3 p}{(2\pi)^3} \frac{(\vec{p})^4}{E_a}
	c^m\,C^{(m)\mn\al\beta}V_{\al \beta}.
	\label{integral_form}
	\eea
	The left-hand side of the eq.~\eqref{Boltzmann_eq} has been written in terms of the projection operators as
	\bea
	-\frac{f^0}{2 T}p^\mu p^\nu V^{\al\beta} \left[\mathscr{P}^{(0)}_{<\mn>\al \beta}+\mathscr{P}^{(1)}_{<\mn>\al \beta}+\mathscr{P}^{(-1)}_{<\mn>\al \beta}+\mathscr{P}^{(2)}_{<\mn>\al \beta}+\mathscr{P}^{(-2)}_{<\mn>\al \beta}\right].
	\label{lhs}
	\eea
	Substituting $\delta f_a$ on the right hand side of eq~\eqref{Boltzmann_eq}, we get 
	\bea
	-\frac{u\cdot p}{\tau_a}\left(1-\frac{qB \tau_a}{u\cdot p} b^{\mn}p_\nu \frac{\partial}{\partial p^\mu}\right)\sum_{m=-2}^{2} c^m C^{(m)}_{\mn\al\beta}\, p^\mu p^\nu V^{\al\beta}.
	\label{rhs}
	\eea
	Equating eq.~\eqref{lhs} and eq.~\eqref{rhs} after writing the fourth rank tensors in terms of second tensors , we get (see Refs.~\cite{Dash:2020vxk,Tuchin:2011jw,Das:2019pqd})
	\bea
	c^0&=& \frac{1}{2T} \frac{f^0 \tau_a}{(u\cdot p)},\nn\\
	c^{1+}&=& \frac{1}{2T} \frac{u\cdot p}{(u\cdot p)^2+(q B\tau_a)^2}f^0 \tau_a,\nn\\
	c^{2+}&=& \frac{1}{2T} \frac{u\cdot p}{(u\cdot p)^2+(2 q B\tau_a)^2}f^0 \tau_a,\nn\\
	c^{1-}&=& \frac{1}{2T} \frac{qB}{(u\cdot p)^2+(q B\tau_a)^2}f^0 \tau_a^2,\nn\\
	c^{2-}&=& \frac{1}{T} \frac{q B}{(u\cdot p)^2+(2 q B\tau_a)^2}f^0 \tau_a^2.
	\eea
	Employing eq.~\eqref{integral_form} we can find  the shear viscosity coefficients as 
	\bea
	\eta_\parallel&=& \frac{2}{15}\sum_a \int \frac{d^3 p}{(2\pi)^3}\frac{|\vec p|^4}{E_a} c^0= \frac{1}{15T}\sum_a \int \frac{d^3 p}{(2\pi)^3}\frac{|\vec p|^4}{E_a^2}f^0 \tau_a ,\\
	\label{visco_para}
	\eta_{\perp}&=& \frac{2}{15}\sum_a \int \frac{d^3 p}{(2\pi)^3}\frac{|\vec p|^4}{E_a} c^{1+}= \frac{1}{15T}\sum_a \int \frac{d^3 p}{(2\pi)^3} \frac{|\vec p|^4}{(u\cdot p)^2+(q B\tau_a)^2}f^0 \tau_a,\\
	\eta_{\perp}'&=& \frac{2}{15}\sum_a \int \frac{d^3 p}{(2\pi)^3}\frac{|\vec p|^4}{E_a} c^{2+}= \frac{1}{15T}\sum_a \int \frac{d^3 p}{(2\pi)^3} \frac{|\vec p|^4}{(u\cdot p)^2+(2q B\tau_a)^2}f^0 \tau_a,\\
	\eta_{\times}&=& \frac{2}{15}\sum_a \int \frac{d^3 p}{(2\pi)^3}\frac{|\vec p|^4}{E_a} c^{1-}= \frac{1}{15T}\sum_a \int \frac{d^3 p}{(2\pi)^3}\frac{|\vec p|^4}{E_a} \frac{qB}{(u\cdot p)^2+(q B\tau_a)^2}f^0 \tau_a^2,\\
	\eta_{\times}'&=& \frac{2}{15} \sum_a\int \frac{d^3 p}{(2\pi)^3}\frac{|\vec p|^4}{E_a} c^{2-}= \frac{2}{15T}\sum_a \int \frac{d^3 p}{(2\pi)^3}\frac{|\vec p|^4}{E_a} \frac{qB}{(u\cdot p)^2+(2q B\tau_a)^2}f^0 \tau_a^2.
	\label{visco_hall}
	\eea
	In the presence of a nonzero magnetic field, the shear stress tensor is written using the available basis (as discussed in this section), having a component parallel to the magnetic field. The subscript $\parallel$ denotes this parallel component. The subscripts $\perp$ and $\times$ are the perpendicular and Hall components. In the absence of a magnetic field, the Hall component is zero, whereas the perpendicular component becomes the same as the parallel component.
	
	From eq.~\ref{sol_KG}, the solution of eq.~\ref{kg_eq} for charged particles looks like $\sim e^{-i P\cdot X_{\tilde{x}}}Exp[-eB/2(x-\frac{p_y}{eB})]H_\nu(x-\frac{p_y}{eB})$. To find out the density of state we consider box of length $[L_1,L_2,L_3]$ and infinite volume limit would be taken at the end. $p_y$ takes the discrete values i.e. $p_y=\frac{2\pi n}{L_2}$ with integers $n$. The particle is localized in $x\sim p_y/eB=\frac{2\pi n}{ eB L_2}$. As $x$ lies in interval $[0,L_1]$, we can write $0<n<\frac{L_1L_2}{2\pi}eB$. So the number of states in transverse area is $\frac{L_1L_2}{2\pi}eB$. Now the number of states per unit volume in $\Delta p_z$ interval becomes $\sim \frac{1}{V}\frac{L_1 L_2 L_3}{2\pi}eB \frac{\Delta p_z}{2\pi}=eB \frac{\Delta p_z}{(2\pi)^2}$. Using this argument, for the case of the charged particles we use the phase space~\cite{Hattori:2016lqx} as
		\bea
		\int \frac{d^3 p}{(2\pi)^3} \longrightarrow \frac{|eB|}{(2\pi)^2}\int dp_z.
		\eea
	For the neutral particle, the Hall components of the viscosity coefficients are zero, and the perpendicular components are equal to the parallel parts.
	
	In the presence of a finite magnetic field, the momentum of the charged particles becomes anisotropic. The energy dispersion relation of the charged scalar particle of charge $q$ gets modified as 
	\bea
	E_n=\sqrt{p_L^2+(2n+1)|qB|+m^2},
	\eea
	where $p_z$ is the momentum of the particle parallel to the direction of the magnetic field, and $m$ is the mass of that particle. Here $n=0,1,2,3...$ denotes the Landau levels. In our case, we consider the magnetic field to be in the z-direction. We also work in a strong magnetic field limit, i.e., the magnetic field is much greater than the temperature square scale. In this approximation, we can safely consider the confinement of charged particles in the lowest Landau level (LLL), and the energy dispersion in LLL is given as
	\bea
	E&=&\sqrt{p_3^2+m^2+|qB|}.
	\eea
	Here, we define the notation $\tilde p^\mu=(E,p^3)$. 
	For the charged pions equilibrium distribution takes the form as $f_a^0=\exp{(-u_\al \tilde p^\al/T)}$.
	Now, the uncharged particles are not affected by the magnetic field, therefore the distribution is given by  $f_a^0=\exp{(-u_\al p^\al/T)}$, where $p^{\mu}=(E, \vec{p})$ with $\om=\sqrt{\vec{p}^2+m^2}$.
	
	\section{Linear Sigma Model}
	\label{sec:LSM}
	The LSM model is a simplistic effective model of pions. Here we use it to calculate transport coefficients.
	In general, the LSM Lagrangian consists of $N$ bosonic fields. For $N=4$, it represents the theory of three pions $(\pi_i)$ and one sigma $(\sigma)$ fields.
	The LSM Lagrangian density~\cite{Chakraborty:2010fr,Scavenius:2000qd} for $N=4$ is
	\bea
	\mathcal{L}&=&\frac{1}{2} (\partial_\mu \sigma)^2+\frac{1}{2} (\partial_\mu \boldsymbol{\pi} )^2-V(\sigma,\boldsymbol{\pi}),
	\eea
	where the potential term
	\bea
	V(\sigma, \boldsymbol{\pi})=\frac{\lambda}{4}(\sigma^2+\boldsymbol{\pi}^2-f^2)^2-H\sigma.
	\eea
	Here $H \sigma$ is the explicit chiral symmetry breaking term that gives the pion mass. The scalar field $\sigma$ takes the vacuum expectation value $v$ as $\sigma=v+\Delta$, where $\Delta$ is the fluctuation and $v$ is determined by the symmetry breaking term as
	\bea
	\lambda v (v^2-f^2)=H.
	\eea
	Other parameters $\lambda$, $H$ and $f$ are expressed in terms of pion decay constant $f_\pi$, pion masses $(m_\pi)$ and sigma masses $(m_\sigma)$ i.e.
	\bea
	\lambda &=&\frac{m_\sigma^2-m_{\pi}^2}{2 f_{\pi}^2},\nn\\
	H&=&f_\pi m_\pi^2,\nn\\
	f^2&=& f_\pi^2 \frac{m_\sigma^2-3 m_\pi^2}{m_\sigma^2-m_\pi^2}.
	\eea
	For calculation we are taking vacuum pion mass $m_\pi=140$ MeV, vacuum $\sigma$ masses $m_\sigma=\{500,700\}$ MeV and  decay constant $f_\pi=93$ MeV. 
	
	We would continue our calculations in the isospin pion basis representing the physical pions. The physical pions can be expressed in terms of Cartesian pion fields as,
	\bea
	\pi^0&=&\pi_3,\\
	\pi^\pm&=& \frac{1}{\sqrt{2}}(\pi_1\pm i \pi_2).
	\eea
	In physical pion basis interaction Lagrangian can be written as
	\bea
	\mathcal{L}_{int}&=& \frac{\lambda}{4}\bigg(\sigma^4+(\pi^0)^4+(\pi^+)^4+(\pi^-)^4+2 (\pi^0)^2 (\pi^+)^2+2 (\pi^0)^2 (\pi^-)^2+2 (\pi^0)^2 \sigma^2\nn\\
	&+&2 (\pi^+)^2 (\pi^-)^2+2 (\pi^+)^2 \sigma^2+2 (\pi^-)^2 \sigma^2+4 v \sigma (\pi^0)^2+4 v \sigma (\pi^+)^2+4 v \sigma (\pi^-)^2+4 v \sigma^3\bigg).
	\label{Lagrangian}
	\eea
	From the above interaction Lagrangian, one can find the probable interactions between the mesons. 
	
	As we are considering a magnetic field in $z$-direction, the magnetic field $\vec{B}=B \hat z$. The covariant four-derivative $D_\mu=\partial_\mu+Q A_\mu$ replaces the four-derivative $\partial_\mu$ in the kinetic terms of the Lagrangian for the charged pions. Here, $Q=e$ for $\pi^\pm$ and $A^\mu=\{0,0,x B, 0\}$.

	\subsection{Thermodynamics}
	The temperature dependence of the effective masses of pions and the condensate $v$ is rigorously discussed in Refs.~\cite{Kadam:2015fza,Ayala:2002qy,Ayala:2000px}. There is a significant difference between meson masses at low temperature, i.e., chiral symmetry is broken, and symmetry is restored at around $245$ MeV. We are also considering only temperature dependence on effective masses in our case.
	\begin{center}
		\begin{figure}[tbh]
			\begin{center}
				\includegraphics[scale=0.42]{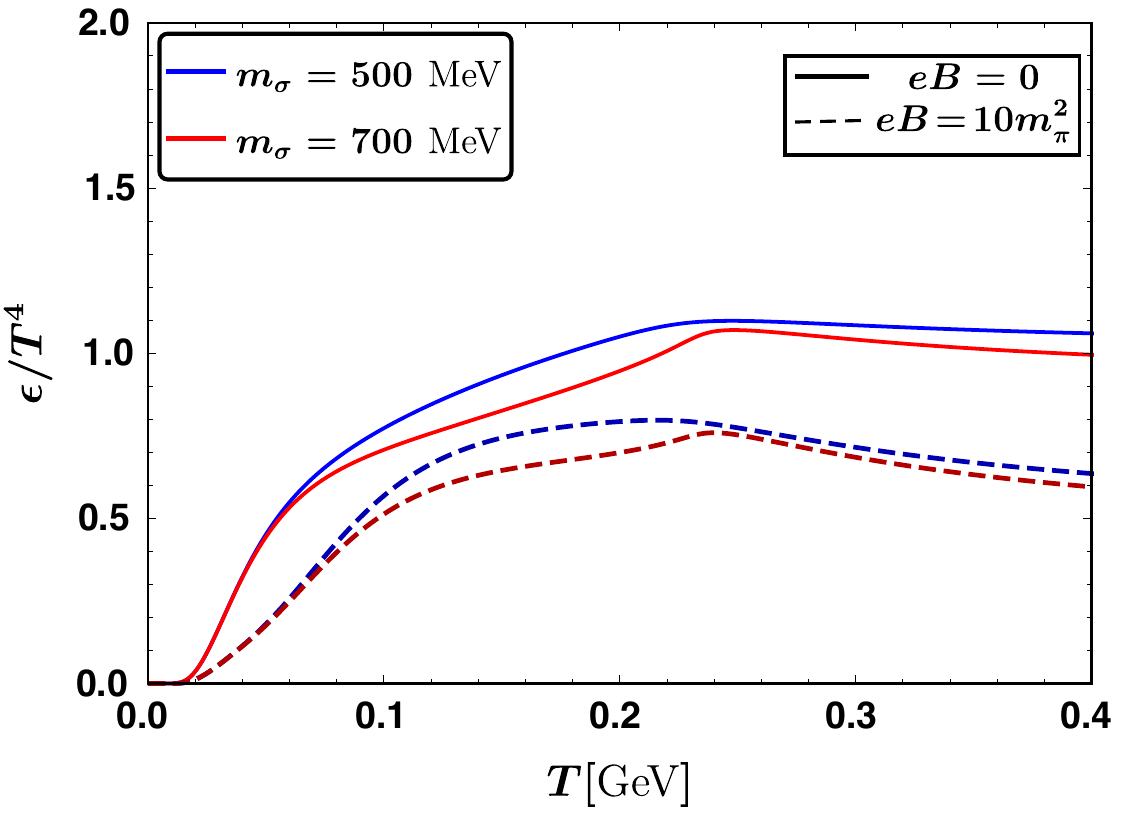}
				\includegraphics[scale=0.42]{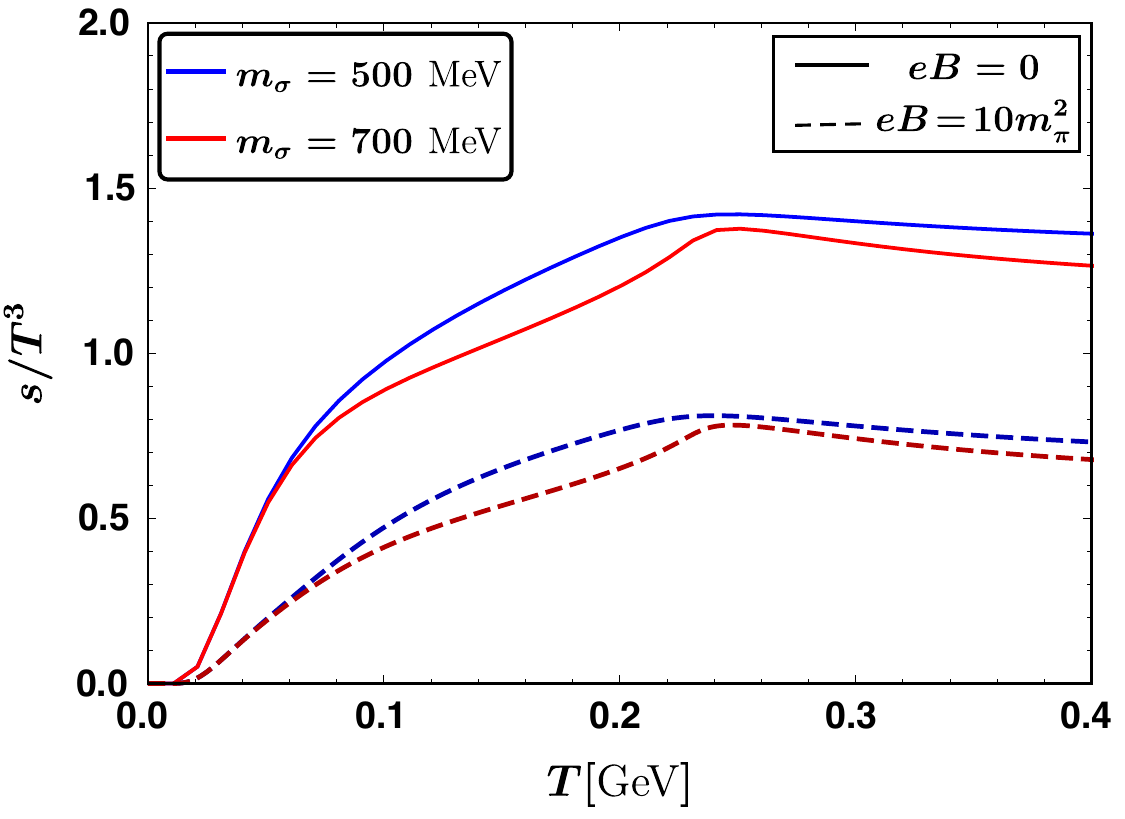}
			\end{center}
			\caption{Energy density (left) and entropy density (right) as functions of temperature
				for three different vacuum sigma masses in presence(dashed line) and absence(solid line) of the magnetic field.}
			\label{energy_6}
		\end{figure}
	\end{center}
	In a strong magnetic field, energy density, entropy density, pressure, and other thermodynamics quantities depend on both temperature and magnetic field effect. The phase factors and energy dispersion is modified for the charged particles. So energy density becomes
	\bea
	\eps_B&=&\sum_{a=\sigma,\pi^0} \int \frac{d^3p}{(2\pi)^3}\,E_a f^0(E_a/T)+\sum_{a=\pi^\pm} \frac{|eB|}{2\pi}
	\int_{-\infty}^{\infty}\frac{dp_z}{2\pi}\,E_a f^0(E_a/T),
	\eea
	where $E_{\sigma}=\sqrt{p^2+\bar m_{\sigma}^2}$, $E_{\pi^0}=\sqrt{p^2+\bar m_{\pi}^2|}$ and $E_{\pi_\pm}=\sqrt{p_z^2+\bar m_{\pi}^2+|eB|}$. Here $\bar m(T)$ is considered as temperature dependent effective mass which comes from the mean field.
	Similarly, the pressure $P$ and the entropy density $s$  can be written as
	\bea
	P_B&=& \sum_{a=\sigma,\pi^0} \int \frac{d^3p}{(2\pi)^3}\,\frac{|\vec p|^2}{E_a} f^0(E_a/T)+\sum_{a=\pi^\pm} \frac{|eB|}{2\pi}
	\int_{-\infty}^{\infty}\frac{dp_z}{2\pi}\,\frac{p_z^2}{E_a} f^0(E_a/T)
	\eea
	and 
	\bea
	s_B&=&\frac{1}{3T^2}\sum_{a=\sigma,\pi^0} \int \frac{d^3p}{(2\pi)^3}\,|\vec p|^2 f^0(E_a/T)+\frac{1}{3T^2}
	\sum_{a=\pi^\pm} \frac{|eB|}{2\pi}\int_{-\infty}^{\infty}\frac{dp_z}{2\pi}\,p_z^2 f^0(E_a/T).
	\eea
	In fig.~\ref{energy_6} energy density $\eps$ scaled with $T^4$ and entropy density $s$ scaled with 
	$T^3$ are plotted for $m_\sigma=500$ MeV and $700$ MeV respectively. The dashed lines represent the values of the corresponding thermodynamic quantities in the presence of the magnetic field. In the presence of a magnetic field, both the energy and entropy density decrease. 

	\subsection{Scattering amplitudes and interaction frequency}
	\label{sec:scattering}
	\subsubsection{Thermal case}
	Here we present the interaction frequency for pure thermal medium. For the pure thermal medium, the  matrix elements are~\cite{Chakraborty:2010fr}
	\bea
	\mathcal{M}_{fi}(\sigma\sigma|\sigma\sigma)&=&-6\lambda-36 \lambda^2 v^2\bigg(\frac{1}{s-m_\sigma^2}+\frac{1}{t-m_\sigma^2}+\frac{1}{u-m_\sigma^2}\bigg),\nn\\
	\mathcal{M}_{fi}(\pi^a\pi^b|\pi^c\pi^d)&=& -2 \lambda \bigg(\frac{s-m_\pi^2}{s-m_\sigma^2}\delta_{ab}\delta_{cd}+\frac{t-m_\pi^2}{t-m_\sigma^2}\delta_{ac}\delta_{bd}+\frac{u-m_\pi^2}{u-m_\sigma^2}\delta_{ad}\delta_{cb}\bigg),\nn\\
	\mathcal{M}_{fi}(\pi\pi|\sigma\sigma)&=&-2\lambda-4 \lambda^2 v^2\bigg(\frac{3}{s-m_\sigma^2}+\frac{1}{t-m_\sigma^2}+\frac{1}{u-m_\sigma^2}\bigg),\nn\\
	\mathcal{M}_{fi}(\pi\sigma|\pi\sigma)&=&-2\lambda-4 \lambda^2 v^2\bigg(\frac{1}{s-m_\sigma^2}+\frac{1}{t-m_\sigma^2}+\frac{1}{u-m_\sigma^2}\bigg).
	\eea
	The poles in the $s$ and $u$ channels cause issues in the scattering amplitudes, resulting in divergent integrals. The divergence can be cured by introducing the thermal width of mesons violating the crossing symmetries. These terms come from the three-point vertices, and we have excluded them in the equation of state. So we are avoiding those terms taking the infinity limits of $s, t,$ and $u$. Finally, we are left with the constant scattering amplitudes, and those are given as,
	\bea
	\mathcal{M}_{fi}(\sigma\sigma|\sigma\sigma)&=&-6\lambda, \\
	\mathcal{M}_{fi}(\pi^a\pi^a|\pi^a\pi^a)&=&-6\lambda ,\,\,\{a=0,+,-\}\\
	\mathcal{M}_{fi}(\pi^+\pi^-|\pi^+\pi^-)&=&-2\lambda, \\
	\mathcal{M}_{fi}(\pi^0\pi^0|\sigma\sigma)&=&-2\lambda, \\
	\mathcal{M}_{fi}(\pi^a\sigma|\pi^a\sigma)&=&-2\lambda ,\,\,\{a=0,+,-\}\\
	\mathcal{M}_{fi}(\pi^b\pi^0|\pi^b\pi^0)&=&-2\lambda ,\,\,\{b=+,-\}.
	\eea
	For $a+b \rightarrow c+d$ type interaction, interaction frequency $\om_a=1/\tau_a$ is written as~\cite{Abhishek:2017pkp}
	\bea
	\omega^a_{\text{th}}(E_a)&\equiv&\tau^{-1}(E_a)=\sum_{bcd}\frac{1}{1+\delta_{ab}}\int \frac{d^3 p_b d^3 p_c d^3 p_d}{(2\pi)^5} 
	\frac{|\mathcal{M}(a b\rightarrow c d)|^2}{16 E_a E_b E_c E_d}\delta^4(p_a+p_b-p_c-p_d)f_b^{eq}.
	\label{om}
	\eea
	In the centre of mass frame, the interaction frequency can be written in simplified form from eq.~\eqref{om} as
	\bea
	\om^a_{\text{th}}&=&\frac{1}{256 \pi^3 E_a} \sum_{bcd}\frac{1}{1+\delta_{ab}}\int_{m_b}^{\infty} dE_b\sqrt{E_b^2-m_b^2} \int_{-1}^{1}\frac{dx}{p_{ab}\sqrt{s}}(t_{\text{max}}-t_{\text{min}})\,\,|\mathcal{M}|^2f^{eq}(E_b),
	\label{om_th}
	\eea
	where
	\bea
	p_{ab}(s)=\frac{1}{2\sqrt{s}}\sqrt{\lambda(s,m_a^2,m_b^2)},
	\eea
	with the kinematic function $\lambda(x,y,z)=x^2+y^2+z^2-2(xy+yz+zx)$. Other quantities are defined as
	\bea
	s&=& 2 E_a E_b \bigg(1+\frac{m_a^2+m_b^2}{2E_a E_b}-\frac{p_a p_b}{E_a E_b}x\bigg),\\
	t_{\text{max,min}}&=& m_a^2+m_c^2-\frac{1}{2s} (s+m_a^2-m_b^2)(s+m_c^2-m_d^2)\pm \frac{1}{2s}\sqrt{\lambda(s,m_a^2,m_b^2)\lambda(s,m_c^2,m_d^2)}.
	\eea

	\subsubsection{In the presence of magnetic field}
	A finite magnetic field would affect the scattering amplitudes containing charged pions as the charged pions interact with the magnetic field. In this section, we calculate the magnetic field-affected interaction rates. Here we also consider the four-point interactions. Starting from the $S$-matrix elements, we end up with the interaction rates of corresponding processes. The calculations of the matrix elements involve the Klein-Gordon solutions of the charged scalar particles in the presence of the magnetic field. The solutions to the Klein-Gordon equation are discussed in Appendix~\ref{KG_sol}. As we are confining ourselves in the strong magnetic field case, we have obtained the interaction rates only for the lowest Landau level.
	
	The $S$-matrix element for $\pi^b(\vec p_{\tilde x},m)\,+\,\pi^b(\vec k_{\tilde x},n) \rightarrow \pi^b(\vec p'_{\tilde x},m')\,+\,\pi^b(\vec k'_{\tilde x},n')$ scattering in presence of magnetic field is written as
	\bea
	S_{fi}&=&4!\frac{\lambda}{4} \int d^4 X \bra{\pi^b(n',\vec k'_{\tilde x}) \pi^b(m',\vec p'_{\tilde x})}(\pi^b)^4\ket{\pi^b(n,\vec k_{\tilde x})\pi^b(m,\vec p_{\tilde x})},\,\,\,\{b=+,-\}\nn\\
	&=& 4!\frac{\lambda}{4} \int d^4 X \frac{e^{-i (P+K-P'-K')\cdot X_{\tilde x}} }{\sqrt{16 E_n E_m E_n' E_m' (L_y L_z)^4}} f_n(x,\vec k_{\tilde x})f_m(x,\vec p_{\tilde x})f_{n'}^*(x,\vec k'_{\tilde x})f_{m'}^*(x,\vec p'_{\tilde x})\nn\\
	&=&(2\pi)^3 \delta^{(3)}_{\tilde x} (p+k-p'-k') \frac{1}{\sqrt{16 E_n E_m E_n' E_m' (L_y L_z)^4}} \mathcal{M}_{fi},
	\label{S1}
	\eea
	where $ \delta^{(3)}_{\tilde x}$ implies the $\delta$-function for all the space-time coordinates except x. In this case four-momentum conservation is not appearing through the delta function as the $x$-component of the momentum is not a good quantum number.
	The matrix element $\mathcal{M}_{fi}$ from eq.~\eqref{S1} can be read as
	\bea
	\mathcal{M}_{fi}(\pi^b \pi^b|\pi^b\pi^b)=4! \frac{\lambda}{4}\int dx  f_n(x,\vec k_{\tilde x})f_m(x,\vec p_{\tilde x})f_{n'}^*(x,\vec k'_{\tilde x})f_{m'}^*(x,\vec p'_{\tilde x}),\,\,\,\{b=+,-\}.
	\eea
	and similarly we can write the scattering amplitudes for $\pi^+(k')+\pi^-(p')\rightarrow \pi^+(k)+\pi^-(p)$ as
	\bea
	\mathcal{M}_{fi}(\pi^+ \pi^-|\pi^+\pi^-)= 2 \lambda\int dx  f_n(x,\vec k_{\tilde x})f_m(x,\vec p_{\tilde x})f_n^*(x,\vec k'_{\tilde x})f_m^*(x,\vec p'_{\tilde x}).
	\eea
	Other scattering amplitudes affected by magnetic fields are 
	\bea
	\mathcal{M}_{fi}(\pi^b \sigma|\pi^b\sigma)&=&2 \lambda\int dx e^{i(k_x-k_x') x} f_n(x,\vec k_{\tilde x})f_m^*(x,\vec p'_{\tilde x}),\,\,\,\{b=+,-\}\\
	\mathcal{M}_{fi}(\pi^b \pi^0|\pi^b\pi^0)&=&2 \lambda\int dx e^{i(k_x-k_x') x} f_n(x,\vec k_{\tilde x})f_m^*(x,\vec p'_{\tilde x}).\,\,\,\{b=+,-\}
	\eea
	As we are considering strong magnetic field, we restrict ourselves to the lowest Landau levels.
	For $\pi^+(k_a)+\pi^+(k_b)\rightarrow \pi^+(k_c)+\pi^+(k_d)$ and $\pi^-(k_a)+\pi^-(k_b)\rightarrow \pi^-(k_c)+\pi^-(k_d)$ scatterings, the interaction frequencies of ``$\pi^b(b=\pm)$" for these processes are given by
	\bea
	\omega^b_1 &=&\frac{1}{2} \int \frac{dk_y^b dk_z^b}{(2\pi)^2}\frac{dk_y^c dk_z^c}{(2\pi)^2}\frac{dk_y^d dk_z^d}{(2\pi)^2}
	(2\pi)^3\delta^{(3)}_{\tilde x}(k_a+k_b-k_c-k_d)\nonumber\\
	&\times&\frac{1}{16 E_a E_b E_c E_d}|\mathcal{M}_{fi}(\pi^+ \pi^+|\pi^+\pi^+)|^2 f_b^{eq},
	\eea
	with
	\bea
	|\mathcal{M}_{fi}|^2&=&(6\lambda)^2N_0^8 \frac{\pi}{2|eB|}\exp\left\{-\frac{(k_y^a+k_y^b+k_y^c+k_y^d)^2-4(k_y^a)^2-4(k_y^b)^2-4(k_y^c)^2-4(k_y^d)^2}{4|eB|}\right\}.
	\label{matrix_elem}
	\eea
	After integration over $k_y^b$, $k_y^c$, $k_y^d$ we get,
	\bea
	\omega^b_1&=& \frac{1}{2}6^2\lambda^2N_0^8\frac{\pi^2}{2}\frac{(2\pi)^3}{16(2\pi)^6}\int dk_z^b\, dk_z^c\, dk_z^d\, \delta(E_a+E_b-E_c-E_d)\delta(k_z^a+k_z^b-k_z^c-k_z^d)
	\frac{1}{E_a E_b E_c E_d}f_b^{eq}\nn\\
	&=&\frac{1}{2}6^2\frac{\lambda^2}{32}\frac{|eB|^2}{\pi^2}\frac{\pi^2}{(2\pi)^3}\int dk_z^b\, dk_z^c\,\delta(E_a+E_b-E_c-E_d)\frac{1}{E_aE_bE_c \sqrt{m^2+(k_z^a+k_z^b-k_z^c)^2+|eB|}}f_b^{eq}\nn\\
	&=& \frac{1}{2}6^2\frac{\lambda^2}{32}\frac{|eB|^2}{(2\pi)^3}\int dk_z^b\, dk_z^c\,\bigg(\frac{k_z^a}{E_a}-\frac{k_z^b}{E_b}\bigg)^{-1}\frac{\delta(k_z^c-k_z^a)+\delta(k_z^c-k_z^b)}{E_aE_bE_c \sqrt{m^2+(k_z^a+k_z^b-k_z^c)^2+|eB|}}f_b^{eq}\nn\\
	&=&\frac{1}{2} 6^2\frac{\lambda^2}{32}\frac{|eB|^2}{(2\pi)^3}\int_{-\infty}^{\infty} dk_z^b\, \frac{2}{E_a^2 E_b^2}\bigg(\frac{k_z^a}{E_a}-\frac{k_z^b}{E_b}\bigg)^{-1}f_b^{eq}.
	\eea
	Here we have used the identity
	\bea
	\delta[g(x)]=\sum_{i}\frac{\delta(x-x_i)}{|g'(x_i)|},
	\eea
	where $x_i$ are the roots of $g(x)$.
	
	Similarly, for $\pi^+(k_a)+\pi^-(k_b)\rightarrow \pi^+(k_c)+\pi^-(k_d)$  type scattering the interaction frequencies of ``$\pi^{b}(b=\pm)$" are given by,
	\bea
	\om^b_2&=&2^2\frac{\lambda^2}{32}\frac{|eB|^2}{(2\pi)^3}\int_{-\infty}^{\infty} dk_z^b\, \frac{2}{E_a^2 E_b^2}\bigg(\frac{k_z^a}{E_a}-\frac{k_z^b}{E_b}\bigg)^{-1}f_b^{eq}.
	\eea
	Now we are considering the other scatterings $\pi^b(p)+\sigma(k) \rightarrow \pi^b(p')+\sigma(k')$ 
	and $\pi^b(p)+\pi^0(k) \rightarrow \pi^b(p')+\pi^0(k')$ with $\{b=+,-\}$. In these cases interaction frequency of $\pi^b$ particle is
	\bea
	\om^b_3(p)&=& \int\frac{d^3k}{(2\pi)^3}\frac{d^3k'}{(2\pi)^3}\frac{d^2p'}{(2\pi)^2}
	\frac{(2\pi)^3 \delta^{(3)}_{\tilde x}(p+k-p'-k')}{16E_kE_pE_{p'}E_{k'}}|\mathcal{M}_{fi}|^2 f^{eq}(E_k),\nn\\
	\eea
	with 
	\bea
	|\mathcal{M}_{fi}|^2&=&(2\lambda)^2N_0^4 \frac{\pi}{|eB|}\exp\left\{-\frac{(p_y-p_y')^2+(k_x-k_x')^2}{2|eB|}\right\}.
	\label{matrix_elem2}
	\eea
	After integration over $p_y'$ and $k_z$ we get,
	\bea
	\om^b_3&=&(2\lambda)^2N_0^4\frac{\pi}{16|eB|}\int\frac{d^3k'}{(2\pi)^5}\frac{d^2 k_\perp dp_z'}{E_pE_{k'}E_{p'}\sqrt{k_\perp^2+(p_z'+k_z'-p_z)^2}}\nn\\
	&\times&\delta(E_p+\sqrt{k_\perp^2+(p_z'+k_z'-p_z)^2}-E_{p'}-E_{k'})e^{-\frac{(k-k')_\perp^2}{2|eB|}}f^{eq}\bigg(\sqrt{k_\perp^2+(p_z'+k_z'-p_z)^2}\bigg)\nn\\
	&=&(2\lambda)^2\frac{|eB|}{\pi} \frac{\pi}{16|eB|}\int\frac{d^3k'}{(2\pi)^5}\frac{d^2 k_\perp dp_z'}{E_pE_k'E_{p'}\sqrt{k_\perp^2+(p_z'+k_z'-p_z)^2}}
	2|E_{p'}+E_{k'}-E_p|e^{-\frac{(k-k')_\perp^2}{2|eB|}}\nn\\
	&\times&\delta(k^2_\perp-(E_{p'}+E_{k'}-E_p)^2+(p_z'+k_z'-p_z)^2)f^{eq}\bigg(\sqrt{k_\perp^2+(p_z'+k_z'-p_z)^2}\bigg).
	\eea
	We perform the integration numerically after completing the $k_\perp $ integration using the delta function.
	
	Next, we are calculating the interaction rate of the neutral scalar particles $\sigma$ and $\pi^0$ from  $\pi^b(p)+\sigma(k) \rightarrow \pi^b(p')+\sigma(k')$ 
	and $\pi^b(p)+\pi^0(k) \rightarrow \pi^b(p')+\pi^0(k')$ scatterings with $\{b=+,-\}$. For these types of  interactions, we can write the expression of interaction frequencies of $\pi^0$ and $\sigma$ as
	\bea
	\om^{\sigma,\pi^0}_4(k)&=& \frac{1}{L_x}\int\frac{d^3k'}{(2\pi)^3}\frac{d^2p}{(2\pi)^2}\frac{d^2p'}{(2\pi)^2}
	\frac{(2\pi)^3 \delta^{(3)}_{\tilde x}(P+K-P'-K')}{16E_kE_pE_{p'}E_{k'}}|\mathcal{M}_{fi}|^2 f^{eq}(E_p),
	\eea
	where  the matrix element is same as in Eq.~\eqref{matrix_elem}.
	After integration over $p_y'$ and $k_z'$ we get,
	\bea
	\om^{\sigma,\pi^0}_4&=&(2\lambda)^2N_0^4\frac{\pi}{16|eB|}\int\frac{d^2k'_\perp}{(2\pi)^4 L_x}\frac{d p_y dp_z\, dp_z'}{E_pE_{k}E_{p'}\sqrt{k_{\perp}'^2+(p_z+k_z-p_z')^2}}\nn\\
	&\times&\delta(E_p+E_k-\sqrt{k_{\perp}'^2+(p_z+k_z-p_z')^2}-E_{p'})e^{-\frac{(k-k')_\perp^2}{2|eB|}}f^{eq}(E_p).
	\eea
	We have performed the integration numerically after using the delta function. 
	Here we have used $\int dp_y=|eB| L_x$.
	
	Associated scattering processes to calculate  $\pi^+$  relaxation time i.e. $\tau_{\pi^+}$ are 
	\bea
	&&\pi^++\pi^a\rightarrow \pi^++\pi^a \,\,\,(a=+,-,0),\nn\\
	&&\pi^++\sigma\rightarrow \pi^++\sigma. 
	\eea
	Total interaction frequency for $\pi^{+}$ is obtained as $\om^{\pi^+}=\om_{1}^{\pi^+}+\om_{2}^{\pi^+}+\om_{3}^{\pi^+}$. The equilibration time $\tau_{\pi^+}$ is given by $\tau_{\pi^+}=1/\om^{\pi^+}$. For the only scalar interaction (for example: $\sigma\sigma \rightarrow \sigma\sigma$) we considered the expression of interaction rate from equation~\eqref{om_th}. In the similar fashion we can calculate the interaction frequencies for other particles. Incorporating the estimated relaxation times in eq.~\eqref{visco_para}-\,eq.~\eqref{visco_hall}, we can obtain the viscosity coefficients.

	\section{Results}
	\label{sec:result}
	\begin{center}
		\begin{figure}[tbh]
			\begin{center}
				\includegraphics[scale=0.42]{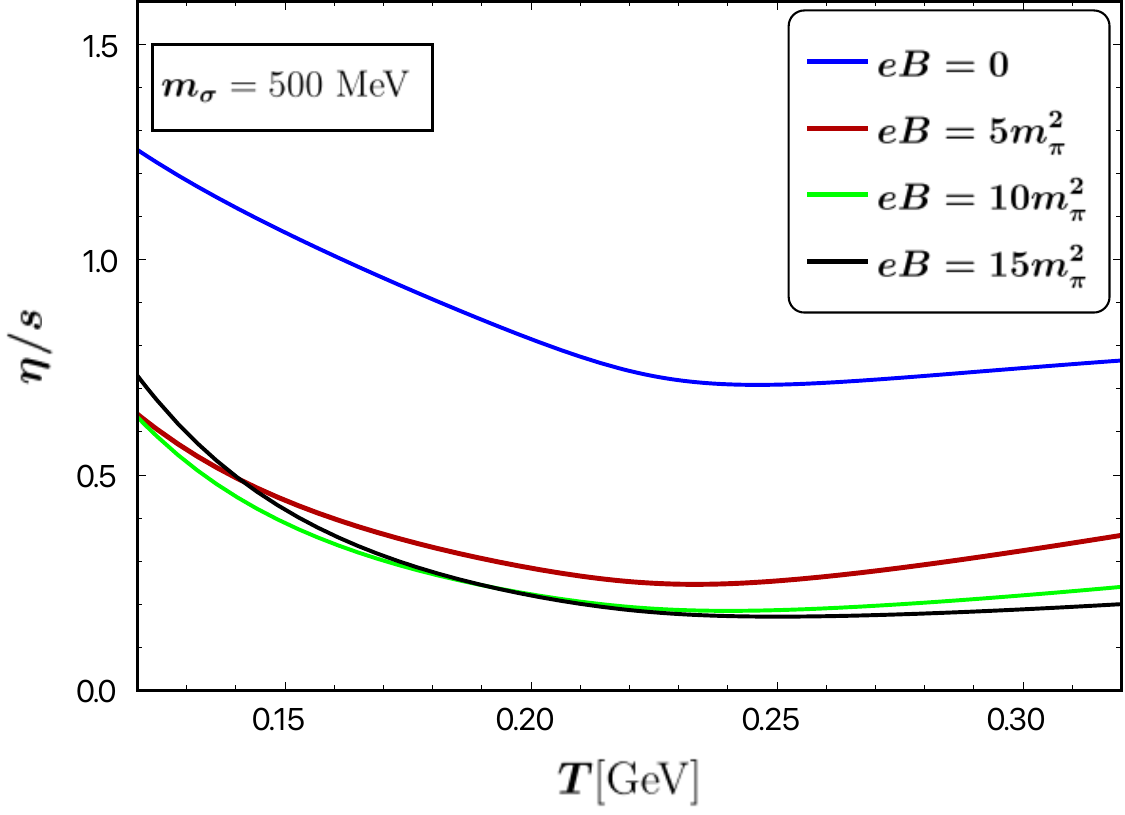}
				\includegraphics[scale=0.42]{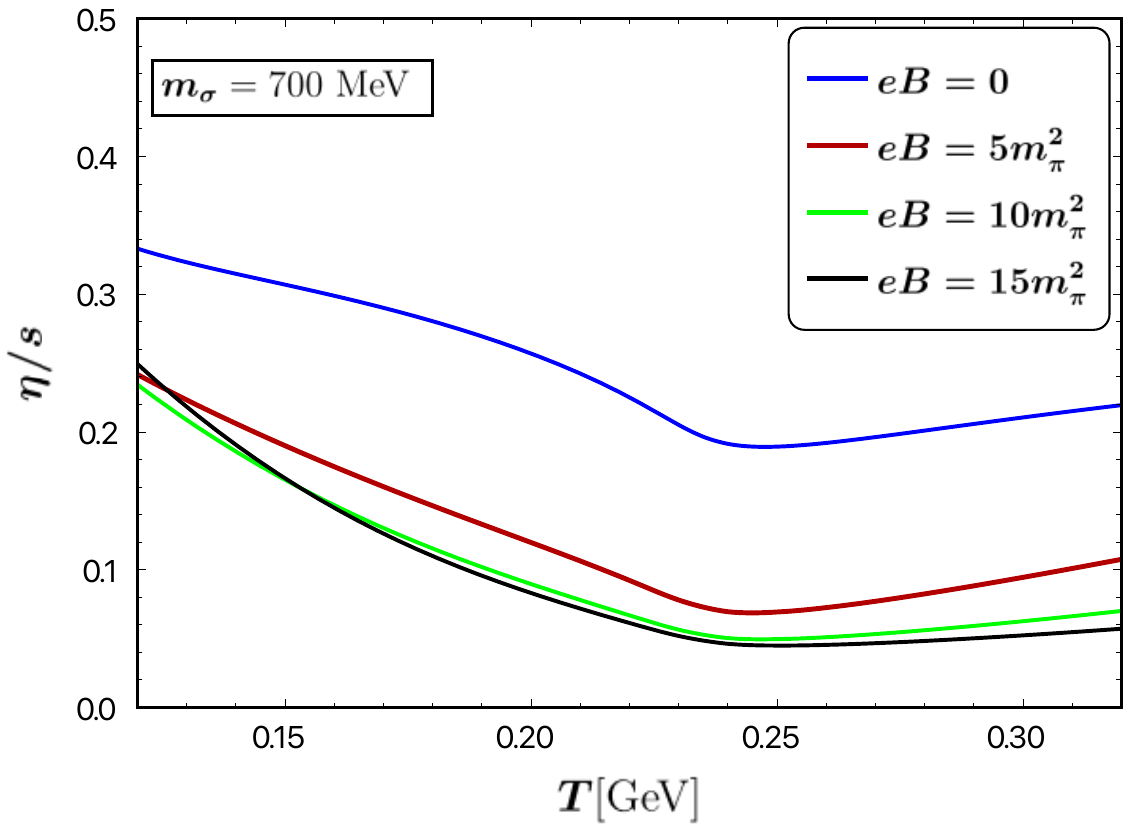}
			\end{center}                  
			\caption{The ratio of shear viscosity to entropy density as a function of temperature for vacuum sigma masses $m_{\sigma}=500\, \text{MeV}$(left) and  $700\, \text{MeV}$(right). In both the plots blue lines indicate the pure thermal case whereas the other lines represent the parallel components of shear viscosity coefficients to entropy ratio for $5 m_\pi^2$ (red line), $10 m_\pi^2$ (green line), $15 m_\pi^2$ (black line).}
			\label{eta}                   
		\end{figure}                   
	\end{center}                    
	%
	We have summarized our results for the anisotropic components of the shear viscosity coefficients in the relaxation time approximation for the nonzero magnetic field. We then briefly discussed the linear sigma model and its thermodynamics. The temperature and magnetic field-dependent nature of the thermodynamic quantities like $s/T^3$ and $\eps/T^3$ are plotted and discussed.
	
	We have revisited the solution of the Klein-Gordon (KG) equation in the presence of a background magnetic field described by a particular vector potential. The quantized nature of the transverse motion of the charged particles emerges to change the particles' energy. The solutions of the KG equation are dependent on the Landau levels. Quantizing the theory, we have calculated the matrix elements using the field operators to obtain the interaction rates. The temperature and magnetic field-dependent interaction rates are incorporated into the thermal relaxation times. In our present study, we have done our evaluation for a strong magnetic field by considering only the lowest Landau level contributions.
	In fig.~\ref{eta} we have compared the pure thermal ($B=0$ case) isotropic viscous coefficient with parallel component of shear viscosity of the thermo-magnetic medium for vacuum sigma masses $m_\sigma=500$ MeV and $700$ MeV. The viscous coefficients are scaled with the entropy density. Only temperature-dependent entropy is considered for the thermal case, whereas temperature $(T)$ and magnetic field $(B)$ dependent entropy is taken for the magnetic case. The plots are shown for three magnetic field strengths, i.e., $5m_\pi^2$ (Redline), $10m_\pi^2$ (Green line), and $15m_\pi^2$ (Blackline). There is a minimum at crossover temperature of $245\, \mbox{MeV}$ for both thermal and magnetic cases. We can also observe that shear viscosity is reduced in the presence of the magnetic field. 
	\begin{center}
		\begin{figure}[tbh!]
			\begin{center}
				\includegraphics[scale=0.43]{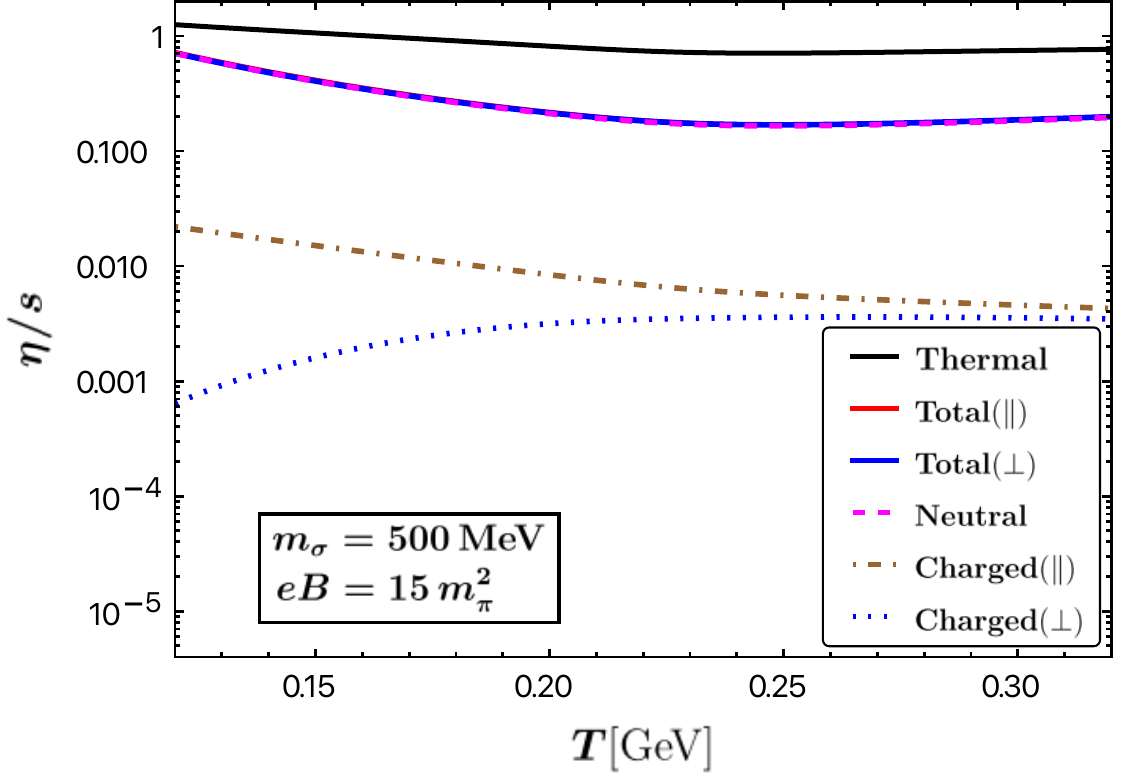}
				\includegraphics[scale=0.51]{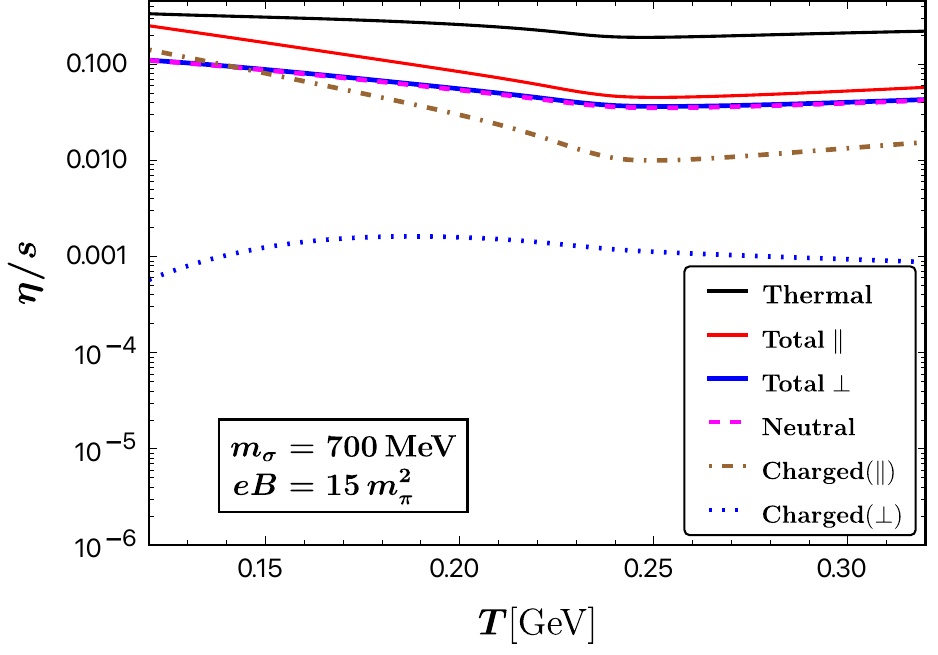}
			\end{center}                  
			\caption{Ratio of parallel $\eta_\parallel$  and perpendicular $\eta_\perp$  shear viscosity components to entropy as a function of temperature $(T)$
				for vacuum sigma mass $m_{\sigma}=500\, \text{MeV}$ (left) and $m_{\sigma}=700\, \text{MeV}$ (right). Magnetic field strength is taken as $15 m_\pi^2$.}
			\label{eta_perp}                   
		\end{figure}                   
	\end{center}          
	Now we will explore the other shear viscous coefficients. LSM has both charged and neutral hadrons. So we studied the perpendicular components for charged and neutral particles differently. As mentioned earlier, neutral particles have a single viscous coefficient, which only contributes to the isotropic shear viscosity. In fig.~\ref{eta_perp}, the solid black line indicates the variation of the scaled isotropic shear viscous coefficient with temperature. The blue dotted line represents the perpendicular component of shear viscosity for the charged particles, whereas the brown dot-dashed line shows the parallel component. The dashed line (magenta) shows the isotropic contribution to the viscous coefficients coming from neutral particles is shown by the dashed line (magenta). Total parallel (solid red) and perpendicular (solid blue) shear viscous coefficients are also plotted to compare with the pure thermal case. The plots are shown for vacuum sigma mass of $500$ MeV (left figure) and $700$ (right figure) MeV. It is observed from the figure that the anisotropic viscous coefficients for the charged particles are quite lower than the neutral hadrons. Note that for $m_\sigma=500$ MeV, total(parallel), total (perpendicular), and neutral hadron contributions coincide as the contribution from the charged hadrons is very small for this case. It is also noted that the Hall type shear viscosity is zero for vanishing baryon chemical potential even in a finite magnetic field.

	\section{Summary and outlook}             
	\label{sec:summary}  
	In the presence of the magnetic field, the charged particles get affected, and the system becomes anisotropic. Therefore, the transport coefficients become anisotropic. This work calculates the shear viscosity of hadronic matter in a strong magnetic field and vanishing chemical potential. We have calculated the parallel and perpendicular components of share viscosity in the relaxation time approximation. We have observed that the shear viscosity to entropy ratio for the neutral hadrons gets modified in the presence of a strong magnetic field because of their interaction with the charged particles. In addition, the shear viscosity for charged hadrons is modified in the thermomagnetic medium. We have observed that the contributions to the total shear viscosity to entropy ratio are more dominant for the charged neutral hadrons than the charged hadrons.
	
	The present investigation is limited to the lowest Landau level (LLL) approximation, and it is valid for a very high value of the magnetic field. To study the effect of the magnetic field on the hadronic transport coefficients at a small to moderate strength of the magnetic field, one should include higher Landau levels. Such calculations are in progress and will be presented elsewhere. 
	\section{Acknowledgment}        
	RG is supported by University Grants Commission (UGC).   N.H. is supported in part by the SERB-MATRICS under Grant No. MTR/2021/000939.                   
	
	\appendix                       
	\section{Charged scalar field}  
	\label{KG_sol}                  
	\subsection{wave function}      
	We consider charged particles in a constant magnetic field. The Klein-Gordon equation becomes~\cite{Greiner:1990tz,Setare:2009lzl}
	\bea                            
	\left(i \frac{\partial }{\partial t}-e A_0\right)^2 \phi(x,y,z,t) =\left((i \vec\nabla+e \vec{A})^2+m^2\right)\phi(x,y,z,t),
	\label{kg_eq}
	\eea
	where the wave function can be written in the following form 
	\bea
	\phi(x,y,z,t)=\phi(x,y,z,) e^{-i E t}.
	\eea
	In our case magnetic field is in $z$-direction i.e. $\vec B=B \hat z$. We choose vector potential as
	\bea
	A^\mu=(0,0,x B,0).\label{vec_pot}
	\eea
	Using the vector potential from Eq.~\eqref{vec_pot}, Eq.~\eqref{kg_eq} becomes
	\bea
	\left(E^2-m^2\right)\phi(x,y,z)=\left(-\nabla^2+2 i e B x \frac{\partial}{\partial y}+e^2 B^2 x^2\right)\phi(x,y,z).
	\eea
	The coordinate $x$ appears through the derivatives, so we expect solution as
	\bea
	\phi(x,y,z)=f(x)e^{i k_y y+k_z z}.
	\eea
	Putting it in above equation we get,
	\bea
	\left(\frac{d^2}{d^2 x}+2 eB x k_y-e^2 B^2 x^2+\eps\right)f(x)&=&0\nn\\
	\left[\frac{d^2 }{d x^2}-(eB x-k_y)^2+(E^2-k_z^2-m^2)\right]f(x)&=&0.
	\eea
	After doing variable transformation i.e.
	\bea
	\xi=\sqrt{|eB|} \left(x-\frac{k_y}{eB}\right),
	\eea
	we arrive to equation
	\bea
	\left(\frac{d^2}{d \xi^2}-\xi^2+a\right)f(x)=0,
	\eea
	where $a=\frac{E^2-k_z^2-m^2}{|eB|}$. The solution of above equation exists when $a=2\nu+1$ for $\nu=0,1,2, \ldots$.
	Energy eigenvalues becomes,
	\bea
	E^2=k_z^2+m^2+(2\nu+1)|eB| ,
	\eea
	and the solution for $f $ is
	\bea
	f_\nu(\xi) \equiv N_\nu e^{-\xi^2/2} H_\nu(\xi),
	\eea
	where $H_\nu$ are Hermite polynomials and normalization constant is 
	\bea
	N_\nu=\left(\frac{\sqrt{|eB|}}{\nu! \,2^\nu \sqrt{\pi}}\right)^{1/2}.
	\eea
	$f_\nu(\xi)$ satisfy the completeness relation 
	\bea
	\sum_n f_n(\xi) f_n(\xi')=\delta(x-x'),
	\eea
	and also
	\bea
	\int_{-\infty}^{\infty} f_\mu(x) f_\nu (x) dx= \sqrt{eB}\, \delta_{\mu, \nu}.
	\eea
	Finally we can write 
	\bea
	\phi_n(x,y,z,t)=e^{-i K\cdot X_{\tilde{x}}} f_n (x,k_y),
	\label{sol_KG}
	\eea
	where $ X_{\tilde{x}}$ is position four vector setting $x$ component to zero. We would use $X$ to represent spacial co-ordinates.
	\subsection{Quantization}
	The scalar field operator can be written in terms of annihilation and creation operator as
	\bea
	\Phi (X)=\sum_{n=0}^\infty \int \frac{d k_y \,dk_z}{2 \pi \sqrt{2 E_n}} \left[e^{-i K\cdot X_{\tilde{x}}} f_n (x,\vec k_{\tilde x}) a(n,\vec k_{\tilde x})+e^{i K\cdot X_{\tilde{x}}} f_n^* (x,\vec k_{\tilde x}) b^\dagger(n,\vec k_{\tilde x})\right].
	\eea
	The field $\Phi(X)$ and $\Pi(X)=\dot\Phi^\dagger$ satisfy the commutation relation
	\bea
	\left[\Phi(X),\Pi(Y)\right]=\delta^{(3)}(\vec x-\vec y).
	\eea
	We can obtain the commutation relation for annihilation and creation operator as
	\bea
	\left[a(n,p_{\tilde x}),a^\dagger(m,p'_{\tilde x})\right]=\delta_{n,m} \delta(k_y-k'_y)\delta(k_z-k'_z),
	\eea
	and similar for $b$ and $b^\dagger$.
	Now we define the one-particle states
	\bea
	\ket{n,\vec k_{\tilde x}}=\frac{2\pi}{\sqrt{L_y L_z}}a^\dagger(n,\vec k_{\tilde x})\ket{0}.
	\eea
	Here we have considered a finite box of sides $(L_x,L_y,L_z)$, which is taken to infinite volume limit at the end.
	The action of field operators on one-particle states reads as
	\bea
	\Phi\ket{\pi^-(n,\vec k_{\tilde x})}&=&\frac{1}{\sqrt{2E_n L_y L_z}}e^{-i K\cdot X_{\tilde x}}f_n(x,\vec k_{\tilde x})\ket{0},\nn\\
	\Phi^\dagger\ket{\pi^+(n,\vec k_{\tilde x})}&=&\frac{1}{\sqrt{2E_n L_y L_z}}e^{-i K\cdot X_{\tilde x}}f_n(x,\vec k_{\tilde x})\ket{0}.
	\eea
	

	%

\end{document}